\documentclass[superscriptaddress,amsmath,amssymb,twocolumn,showpacs]{revtex4-1}

\usepackage{color}
\usepackage{graphicx}

\begin{document}


\title{
The virtual-state character of the $^9$Be $1/2^+$ state in the $^9$Be($\gamma$,$n$)$^8$Be reaction
}


\author{Myagmarjav Odsuren}
\affiliation{Nuclear Research Center, National University of Mongolia, Ulaanbaatar 210646, Mongolia}
\email[]{odsurenn@gmail.com}
\author{Yuma Kikuchi}
\affiliation{RIKEN Nishina Center, Wako 351-0198, Japan}
\author{Takayuki Myo}
\affiliation{General Education, Faculty of Engineering, Osaka Institute of Technology, Osaka 535-8585, Japan}
\affiliation{Research Center for Nuclear Physics (RCNP), Osaka University, Ibaraki 567-0047, Japan}
\author{Masayuki Aikawa}
\affiliation{Nuclear Reaction Data Centre, Faculty of Science, Hokkaido University, Sapporo 060-0810, Japan}
\author{Kiyoshi Kat\=o}
\affiliation{Nuclear Reaction Data Centre, Faculty of Science, Hokkaido University, Sapporo 060-0810, Japan}



\date{\today}

\begin{abstract}
We study the character of the first excited $1/2^+$ state of $^9$Be, which is observed as a low-lying sharp peak in the cross section of $^9$Be($\gamma$,$n$)$2\alpha$ just above the $^8$Be~+~$n$ threshold.
Using the $\alpha$~+~$\alpha$~+~$n$ three-body model, we describe the ground and excited unbound states of $^9$Be above the $\alpha$~+~$\alpha$~+~$n$ threshold.
Applying the complex scaling method to the three-body model, we find no $1/2^+$ resonant solutions with the scaling angle of $\theta \le 15$ degrees, while the low-lying peak in the photodisintegration cross section is reproduced in the present calculation.
It is found that the low-lying peak is dominantly explained by the $^8$Be~+~$n$ component.
Furthermore, using the analytical continuation of the coupling constant of the three-body interaction for the $\alpha$~+~$\alpha$~+~$n$ system, we discuss the virtual-state character of the $1/2^+$ state. 
\end{abstract}


\maketitle

{\it Introduction.}
The neutron capture $(n,\gamma)$ and its inverse $(\gamma,n)$ reactions are interesting topics in nuclear physics since these reactions play important roles in nuclear astrophysics. 
In neutron capture reactions, the $\alpha (\alpha n,\gamma)^9$Be reaction is one of the most interesting ones.
It has been suggested that the reaction rate of this reaction is crucial to understand the productions of heavy elements in supernova explosions~\cite{Terasawa01,Sasaqui05}.

To determine the reaction rate of the $\alpha(\alpha n,\gamma)^9$Be, the $^9$Be$(\gamma,n)^8$Be reaction has been performed by several experiments~\cite{Gibbons59,John62,Fujishiro82,Utsunomiya00,Burda10,Arnold12}.
In the $\alpha(\alpha n,\gamma)^9$Be reaction, a sequential process, $\alpha(\alpha,\gamma)^8$Be$(n,\gamma)^9$Be, has been considered as a dominant one.
However, owing to the short life-time of the $^8$Be ground state ($\sim 10^{-16}$s), a direct measurement of the $^8$Be$(n,\gamma)^9$Be reaction is impossible.
For an alternative way, the cross section of its inverse reaction, $^9$Be$(\gamma,n)^8$Be, has been measured to deduce the cross section of $^8$Be$(n,\gamma)^9$Be.

In the experiments, there is an inconsistency among the observed cross sections of $^8$Be$(n,\gamma)^9$Be in particular the peak just above the $^8$Be~+~$n$ threshold~\cite{Gibbons59,John62,Fujishiro82,Utsunomiya00,Burda10,Arnold12}.
The main difference among the cross sections comes from the contribution from the first excited $1/2^+$ state of $^9$Be via the $E1$ transition. 
This low-lying $1/2^+$ state can have an impact on the reaction rate of $^8$Be$(n,\gamma)^9$Be in stellar environments and supernova explosions~\cite{Terasawa01,Sasaqui05}.
For theoretical side, it is interesting problem to answer how the low-lying $1/2^+$ state of $^9$Be contributes to the $^8$Be$(n,\gamma)^9$Be reaction.

Theoretically, the structure and the reaction mechanism of $^9$Be have been studied by using the $\alpha$~+~$\alpha$~+~$n$ three-body models~\cite{Efros99,Arai03,Garrido10,A_Rodriguez10,Garrido11,Casal14}.
In previous works, the structure of the first excited $1/2^+$ state has been discussed.
Efros {\it et al.}~\cite{Efros99} discussed the structure of $1/2^+$ state from the scattering length for the $^8$Be~+~$n$ scattering.
They show that the scattering length is obtained as a large negative value, which characterizes the first excited $1/2^+$ state as a virtual state.
Also, Arai {\it et al.}~\cite{Arai03} show that the virtual-state character of the $1/2^+$ state from the $R$-matrix analysis.
On the other hand, in Refs.~\cite{Garrido10,A_Rodriguez10}, the first excited $1/2^+$ state is discussed as a three-body resonance of $\alpha$~+~$\alpha$~+~$n$.
In these calculations, they show the importance of a strong mixture of the $^5$He~+~$\alpha$ configuration at the internal region of the $1/2^+$ state.
There exists the contradictions among the theories and the structure of the first $1/2^+$ is still unclear.
Additionally, in Ref.~\cite{Garrido11,Casal14}, the importance of the direct three-body capture for $^9$Be below the $^8$Be~+~$n$ threshold is suggested  recently.
It is important to investigate the relation between the structure of the first excited $1/2^+$ state and the mechanism of the capture reaction.
Comprehensive understanding for the structure of $1/2^+$ of $^9$Be is now required.

The purpose of this work is to understand the structure of the low-lying $1/2^+$ state of $^9$Be and its contribution to the $^9$Be$(\gamma,n)^8$Be reaction.
We describe $^9$Be by using the $\alpha$~+~$\alpha$~+~$n$ three-body model.
To treat the photodisintegration into unbound excited states of $^9$Be, we apply the complex scaling method (CSM)~\cite{Aguilar71,Balslev71,Ho83, Moiseyev98, Aoyama06,Myo14,Odsuren14} to the $\alpha$~+~$\alpha$~+~$n$ three-body model.
The CSM is a powerful tool to investigate many-body resonances and has been extensively used to discuss their structures~\cite{Myo11b,Myo12,Myo14}.
It has been shown that the CSM enables us to describe the resonances from two- to five-body system successfully~\cite{Aoyama06,Myo11b,Myo12,Myo14,Odsuren14}.
Furthermore, the CSM has been developed to describe the breakups of unstable nuclei. 
The CSM has been applied to the Coulomb breakup reactions of two-neutron halo nuclei into core~+~$n$~+~$n$ scattering states and well reproduces the observed cross sections~\cite{Myo01,Myo08,Kikuchi13a}.
From the above, the CSM is promising to discuss the excited states of $^9$Be and the photodisintegration into unbound excited states of $^9$Be on the same footing.

Generally, the resonances are obtained as the pole of $S$-matrix with complex energy eigenvalues in the fourth quadrant of the complex energy plane.
On the other hand, the virtual state is obtained on the negative energy axis of the second Riemann sheet.
In the CSM, we can obtain the solution of the resonances but not of the virtual state.
Such a pole of the virtual state has been discussed by using an analytical continuation method within the two-body system~\cite{Arai03}.
The properties of the virtual state of a three-body system such as $^9$Be is still unclear.
In this work, we investigate not only the eigenvalue distribution but also the photodisintegration cross section in order to clarify the structure of the $1/2^+$ state whether the resonance or the virtual state.

{\it Formalism.}
We solve the Schr\"odingier equation for the $\alpha$~+~$\alpha$~+~$n$ system using the complex-scaled orthogonality condition model~\cite{Saito77}.
The complex-scaled Schr\"dingier equation is given as
\begin{equation}
\hat{H}^\theta \Psi^\nu_{J} (\theta)=E^\theta_\nu\Psi^\nu_{J}(\theta),
\label{eq:cssch}
\end{equation}
where $J$ is the total spin of the $\alpha$~+~$\alpha$~+~$n$ system and $\nu$ is the state index.
The complex-scaled Hamiltonian and wave function are given as
\begin{equation}
\hat{H}^\theta = U(\theta) \hat{H} U^{-1}(\theta)\ \ \ \text{and} \ \ \  \Psi^\nu_J (\theta) = U(\theta) \Psi^\nu_J,
\end{equation}
respectively.
The complex scaling operator $U(\theta)$ transforms the relative coordinate $\boldsymbol{\xi}$ as
\begin{equation}
U(\theta): \boldsymbol{\xi} \rightarrow \boldsymbol{\xi} e^{i\theta},
\end{equation}
where $\theta$ is the scaling angle being a positive real number.

The Hamiltonian for the relative motion of the $\alpha$~+~$\alpha$~+~$n$ three-body system for $^9$Be is given as
\begin{equation}
\hat{H}=\sum^{3}_{i=1}t_{i}-T_\text{c.m.}+\sum^{2}_{i=1}V_{\alpha n}(\boldsymbol{\xi}_{i})+V_{\alpha\alpha}+V_\text{PF}+V_3,
 \label{ew:ham}
\end{equation}
where $t_i$ and $T_\text{c.m.}$ are kinetic operators for each particle and the center-of-mass of the system, respectively.
The interactions between the neutron and the $i$-th $\alpha$ particle is given as $V_{\alpha n}(\boldsymbol{\xi}_i)$, where $\boldsymbol{\xi}_i$ is the relative coordinate between them.
We here employ the KKNN potential~\cite{Kanada79} for $V_{\alpha n}$.
For the $\alpha$-$\alpha$ interaction $V_{\alpha \alpha}$ we employ a folding potential of the effective $NN$ interaction~\cite{Schmid61} and the Coulomb interaction:
\begin{equation}
V_{\alpha\alpha} (r) = v_0\exp{\left(-a r^2\right)} + \frac{4e^2}{r}\text{erf}\left(\beta r\right),
\end{equation}
where $v_0 = -106.09$ MeV, $a = 0.2009$ fm$^{-2}$, and $\beta = 0.5972$ fm$^{-1}$. 
The pseudo potential $V_\text{PF}= \lambda|\Phi_{PF}\rangle\langle\Phi_{PF}|$ is the projection operator to remove the Pauli forbidden states from the relative motions of $\alpha$-$\alpha$ and $\alpha$-$n$~\cite{Kukulin86}.
The Pauli forbidden state is defined as the harmonic oscillator wave functions by assuming the $(0s)^4$ configuration whose oscillator length is fixed to reproduce the observed charge radius of the $\alpha$ particle.
In the present calculation, $\lambda$ is taken as $10^6$ MeV.

To discuss the photodisintegration of $^9$Be, it is important to reproduce the breakup threshold into the $\alpha$~+~$\alpha$~+~$n$.
In the present calculation, we introduces the $\alpha$~+~$\alpha$~+~$n$ three-body potential $V_3$ to reproduce the binding energy of the $^9$Be ground state, $E_\text{gs}$, measured from the $\alpha$~+~$\alpha$~+~$n$ threshold.
The explicit form of $V_3$ is given as
\begin{equation}
V_3 = v_3 \exp{(- \mu \rho^2)},
\label{eq:3bp}
\end{equation}
where $\rho$ is the hyper-radius of the $\alpha$~+~$\alpha$~+~$n$ system.

We solve the eigenvalue problem given in Eq.~(\ref{eq:cssch}) with Gaussian expansion method~\cite{hiyama03}, and obtain the energy eigenvalues and eigenstates (their biorthogonal states) as $\{E^\theta_\nu\}$ and $\{\Psi^\nu_J(\theta)\}$ ($\{\tilde{\Psi}^\nu_J(\theta)\}$), respectively.
Using them, we define the complex-scaled Green's function $\mathcal{G}^\theta(E; \boldsymbol{\xi}, \boldsymbol{\xi}')$ as
\begin{equation}
\mathcal{G}^\theta(E; \boldsymbol{\xi}, \boldsymbol{\xi}')
= \left\langle \boldsymbol{\xi} \left| \frac{1}{E-H^\theta} \right| \boldsymbol{\xi}' \right\rangle
= \sum_\nu\hspace{-0.46cm}\int \frac{\Psi^\nu(\theta) \tilde{\Psi}^\nu(\theta)}{E-E^\theta_\nu}.
\label{eq:CSGF}
\end{equation}
In the derivation of the right-hand side of Eq.~(\ref{eq:CSGF}), we use the extended completeness relation, whose detailed explanation is given in Ref.~\cite{Myo97}.
It is noticed that we take into account outgoing boundary conditions for all open channels of a three-body system in the form of complex energy eigenvalues $E^\theta_\nu$.
The complex-scaled Green's function in Eq.~(\ref{eq:CSGF}) enables us to describe the scattering observables for many-body systems, such as the photodisintegration cross section.

We calculate the cross section of the photodisintegration of $^{9}\textrm{Be}(3/2^{-})~+~\gamma\rightarrow \alpha~+~\alpha~+~n$ in terms of the multipole response.
The cross section is expressed as the following form;
\begin{equation}
\sigma_{E \lambda}^{\gamma}(E_{\gamma})=\frac{(2\pi)^{2}(\lambda+1)}{\lambda [(2\lambda+1)!!]^{2}}\left(\frac{E_{\gamma}}{\hbar c}\right)^{2\lambda-1}\frac{dB(E\lambda,E_{\gamma})}{dE_{\gamma}},
\label{eq:phCS}
\end{equation}
where $E_{\gamma}$ is the incident photon energy and $B(E\lambda)$ is electric transition strength with the rank $\lambda$.
We here calculate the photodisintegration cross section from the ground $3/2^-$ state to $1/2^+$ states in $^9$Be and consider only the $E1$ transition here.
Using the CSM and the complex-scaled Green's function in Eq.~(\ref{eq:CSGF}), the $E1$ transition strength is given as
\begin{equation}
\begin{split}
\frac{dB(E1,E_\gamma)}{dE_\gamma} =& \frac{1}{2J_\text{gs}+1} \sum_\nu\hspace{-0.46cm}\int
\left\langle \tilde{\Psi}_\text{gs} || (\hat{O}^\theta)^\dagger(E1) || \Psi^\nu_{1/2^+}(\theta) \right\rangle \\
&\times\frac{1}{E-E^\theta_\nu}
\left\langle \tilde{\Psi}^\nu_{1/2^+}(\theta) || \hat{O}^\theta(E1) || \Psi_\text{gs} \right\rangle,
\end{split}
\label{eq:E1}
\end{equation}
where $J_\text{gs}$ and $\Psi_\text{gs}$ represent the total spin and the wave function of the ground state, respectively.
The energy $E$ is related to $E_\gamma$ as $E=E_\gamma-E_\text{gs}$. 
From the Eqs.~(\ref{eq:phCS}) and (\ref{eq:E1}), we finally obtain the photodisintegration cross section as
\begin{equation}
\sigma_{E1}^{\gamma}(E_{\gamma})=\frac{16 \pi^{3}}{9\hbar c}E_{\gamma}\frac{dB(E1,E_{\gamma})}{dE_{\gamma}}.
\label{eq:csdef}
\end{equation}

{\it Results.}
We first show the calculated ground-state properties of $^9$Be.
In TABLE~\ref{ta:gspro}, the calculated binding energy and charge and matter radii are listed.
Without the three-body potential in Eq.~(\ref{eq:3bp}), the binding energy of $^9$Be ground state is overbound and the charge radius is slightly small compared to experiments.
To reproduce these quantities, we need the repulsive three-body potential whose parameters are given as $v_3 = 1.10$ MeV and $\mu = 0.02$ fm$^{-2}$.
As results, we reproduce the binding energy and charge radius of $^9$Be ground state simultaneously, while the matter radius is slightly larger than the observed one.
We also confirm that no resonance of the $1/2^+$ state is found with the three-body potential in TABLE~\ref{ta:gspro} for the $\theta = 15$ degrees case.
\begin{table}[tb]
\caption{\label{ta:gspro}
Ground-state properties in comparison with experiments.
The calculated binding energies ($E_\text{gs}$, unit in MeV), charge radii ($R_\text{ch}$, unit in fm), and matter ones ($R_\text{m}$, unit in fm) with and without the three-body potential $V_3$ are listed.}
\begin{ruledtabular}
\begin{tabular}{cccccc}
 & $E_\text{gs}$ & $R_\text{ch}$ & $R_\text{m}$& $v_3$ & $\mu$ \\
\hline
with $V_3$ & 1.57 & 2.53 & 2.42 & 1.10 & 0.02 \\
without $V_3$ & 2.14 & 2.50 & 2.39 & - & - \\
Exp. & 1.5736\footnote{Reference \cite{Tilley04}} & 2.519$\pm$0.012\footnote{Reference \cite{Nortershauser09}} & 2.38$\pm$0.01\footnote{Reference \cite{Tanihata88}} &  &  \\
\end{tabular}
\end{ruledtabular}
\end{table}

\begin{figure}[tb]
\centering{\includegraphics[clip,width=7.5cm]{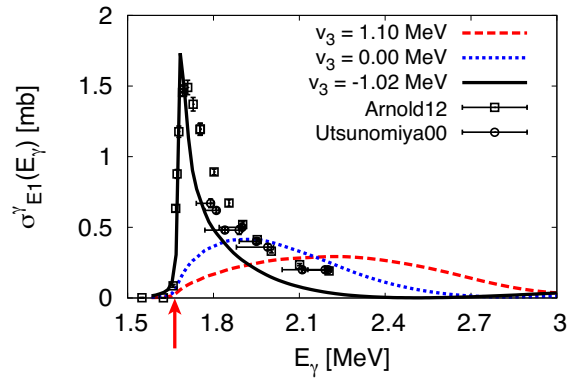}}
\caption{\label{fig:phcs}
Calculated photodisintegration cross sections in comparison with the experimental data.
The red (dashed) and blue (dotted) lines are results with and without the three-body potential, where $v_3 = 1.10$ MeV and $\mu = 0.02$ fm.
The black (solid) line represents the cross section calculated by using an attractive three-body potential with $v_3 = -1.02$ MeV.
The experimental data below $E_\gamma = 2.2$ MeV are taken from Refs.~\cite{Utsunomiya00} and \cite{Arnold12}. The arrow indicates the threshold energy of the $^8$Be(0$^+$)~+~$n$ channel.
}
\end{figure}

Next we discuss the photodisintegration cross section to $1/2^+$ states.
In the present calculation, we fix the ground-state wave function obtained with the three-body potential in TABLE~\ref{ta:gspro}.
In FIG.~\ref{fig:phcs}, we show the calculated cross sections using Eq.~(\ref{eq:csdef}) in comparison with the two sets of the observed data which commonly have peaks just above the $^8$Be~+~$n$ threshold.
The red (dashed) and blue (dotted) lines show the cross sections with and without the three-body potential for excited $1/2^+$ states, respectively, whose parameters are same as those in TABLE~\ref{ta:gspro}.
In both results, the calculated cross sections underestimate the low-lying peak above the $^8$Be~+~$n$ threshold. 

To discuss the observed sharp peak just above the $^8$Be~+~$n$ threshold in the photodisintegration cross section, we change the strength, $v_3$, for the $1/2^+$ state to fit the observed data but its range is fixed as the same as used in the ground state.
We here take the strength as $v_3 = -1.02$ MeV for the $1/2^+$ state and obtain the cross section as shown as the black (solid) line in FIG.~\ref{fig:phcs}.
Our result reproduces the observed peak by using the attractive three-body potential.
We confirm that the calculated cross section rapidly increases just above the $^8$Be~+~$n$ threshold and there is negligibly small strength below this threshold.
We also find that the calculated cross sections show the strong dependence on the strengths of the three-body potentials as shown in FIG.~\ref{fig:phcs}.
This result is interesting and suggests the existence of the three-body unbound state of $^9$Be($1/2^+$), such as a resonance or virtual state.
In relation to the cross section, we discuss the character of the $1/2^+$ state. 

\begin{figure}[tb]
\centering{\includegraphics[clip,width=8cm]{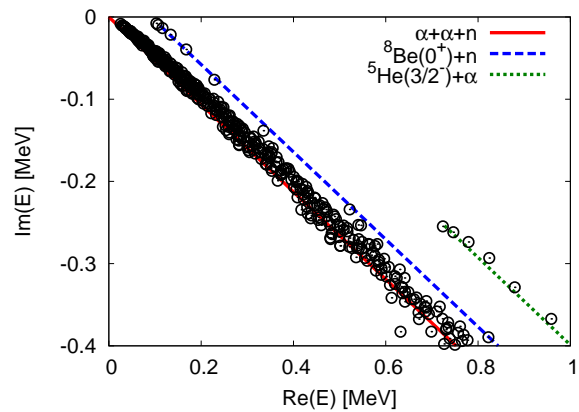}}
\caption{\label{fig:edis}
Distribution of energy eigenvalues of $1/2^+$ states measured from the $\alpha$~+~$\alpha$~+~$n$ threshold with scaling angle $\theta = 15$ degrees.
We here employ the three-body potential with $v_3=-1.02$ MeV and $\mu=0.02$ fm$^{-2}$.
The red (solid), blue (dashed), and green (dotted) lines represent the branch cuts for $\alpha$~+~$\alpha$~+~$n$, $^8$Be(0$^+$)~+~$n$, and $^5$He(3/2$^-$)~+~$\alpha$ continua, respectively. 
}
\end{figure}
We investigate the origin of the low-lying peak above the $^8$Be~+~$n$ threshold more in detial.
For this purpose, we show the distribution of the energy eigenvalues of the $1/2^+$ states by using the CSM.
In the CSM, continuum states are obtained along the branch cuts which start from the threshold energies and are rotated down by 2$\theta$. 
A resonance is obtained as a solution with complex energy of $E^\theta = E_r - i\Gamma/2$ isolated from the continuum ones .
On the other hand, the virtual states and broad resonances, which are located on the second Riemann sheet covered by the rotated first Riemann sheet, cannot be obtained as the isolated pole in CSM.
The contributions from these states to the cross section are scattered into the continuum states rotated on the $2\theta$ lines.
In FIG.~\ref{fig:edis}, we show the distribution of the energy eigenvalues for the $1/2^+$ states calculated with $v_3 = -1.02$ MeV, which reproduces the observed peak as shown in FIG.~\ref{fig:phcs}.
In the present calculation, we find no resonances in the energy eigenvalue distribution.
All energy eigenvalues are located on the 2$\theta$-lines, corresponding to the branch cuts for the $\alpha$~+~$\alpha$~+~$n$, $^8$Be(0$^+$)~+~$n$, and $^5$He(3/2$^-$)~+~$\alpha$ continuum states.

\begin{figure}[tb]
\centering{\includegraphics[clip,width=7.5cm]{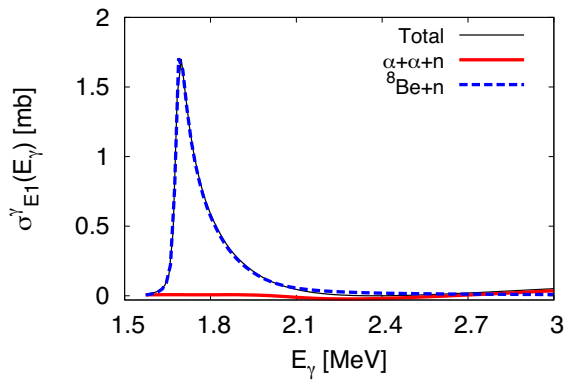}}
\caption{\label{fig:deccs}
Decomposed photodisintegration cross sections.
The red (solid) and blue (dashed) lines are contributions of the $\alpha$~+~$\alpha$~+~$n$ and $^8$Be~+~$n$ continuum states.
The black thin line is same as that in FIG.~\ref{fig:phcs}.
}
\end{figure}

We investigate the contributions of two- and three-body continuum states to the cross section to understand the mechanism of the photodisintegration.
Using the CSM, we decompose the cross section into $^8$Be~+~$n$ and $\alpha$~+~$\alpha$~+$n$ components as shown in FIG.~\ref{fig:deccs}, to clarify the dominant contribution in the cross section~\cite{Myo01}. 
The $^5$He~+~$\alpha$ contribution is found to be negligible in the low-lying region, and we do not show it in FIG.~\ref{fig:deccs}.
From the results, we see that the $^8$Be~+~$n$ component is almost identical to the total cross section.
This fact indicates that the $^8$Be~+~$n$ decay is dominant in the photodintegration.
This decay mode should be related to the structure of the $1/2^+$ state of $^9$Be.

\begin{figure}[tb]
\centering{\includegraphics[clip,width=6.5cm]{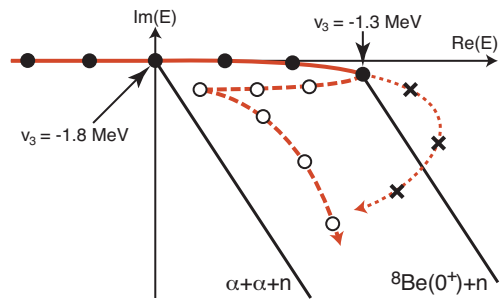}}
\caption{\label{fig:etra}
Pole trajectory of the $^9$Be $1/2^+$ state in complex energy plane by changing the three-body potential.
The closed circles represent the poles obtained as isolated three-body resonances in CSM.
The open circles and the crosses are speculated pole positions for the virtual states and broad resonances, respectively.
}
\end{figure}
To investigate the structure of the $1/2^+$ state of $^9$Be, we calculate the energy eigenvalues of the $\alpha$~+~$\alpha$~+~$n$ system by changing the strength of the three-body potential, $v_3$, which is shown in FIG.~\ref{fig:etra}.
In the present calculation, when the strength of the three-body potential $v_3 = -1.3$ MeV, the resonance pole suddenly appears just below the $^8$Be(0$^+$)~+~$n$ threshold.
This resonance pole with a narrow decay width moves smoothly to the bound state region as the three-body potential becomes more attractive, and we finally obtain the $^9$Be bound state with the region of $v_3 < -1.8$ MeV.
On the other hand, we consider the pole trajectory in the opposite case of the three-body potential with $v_3 > -1.3$ MeV.
If the resonance exists, the pole with a narrow decay width should appear above the $^8$Be~+~$n$ threshold as the analytical continuation from the resonance pole as shown with the crosses in FIG.~\ref{fig:etra}.
However, we found that no resonances appear above the $^8$Be(0$^+$)~+~$n$ threshold for $v_3 > -1.3$ MeV of the three-body potential.
These facts in the pole trajectory show the possibility of the virtual state of the $1/2^+$ state consisting of $^8$Be($0^+$)~+~$n$ when we take $v_3=-1.02$ MeV, which reproduces the experimental cross section.
The existence of the virtual state is consistent with the dominant decay into $^8$Be~+~$n$ in the photodisintegration of $^9$Be.

The virtual state is often discussed in unstable nuclei, such as $^{10}$Li and $^{11}$Li~\cite{Masui00,Myo08,Kikuchi13a}.
For $^{10}$Li, Masui {\it et al.}~\cite{Masui00} showed the explicit pole trajectory of the virtual state of $^{10}$Li as $^9$Li~+~$n$ picture, which is similar to the present $^8$Be($0^+$)~+~$n$ situation.
In FIG~\ref{fig:etra}, we schematically plot the pole trajectory in the region of $v_3 > -1.3$ MeV, considering the analogy with the $^{10}$Li case as shown in FIG.~3(b) of \cite{Masui00}.
This trajectory is located on the second Riemann sheet of the $^8$Be~+~$n$ system which cannot be obtained in the CSM.

\begin{figure}[tb]
\centering{\includegraphics[clip,width=7.5cm]{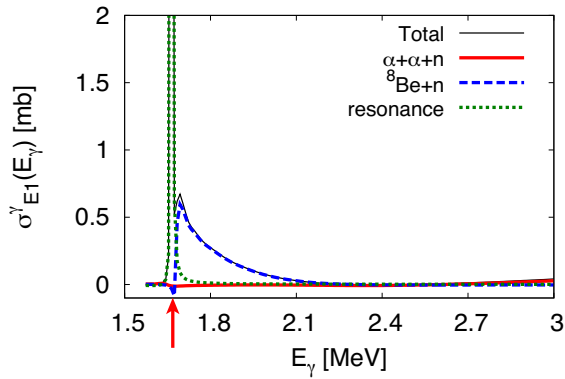}}
\caption{\label{fig:deccs_res}
Decomposed photodisintegration cross sections by using the three-body potential with $v_3=-1.3$.
The red (solid), blue (dashed) lines are contributions of the $\alpha$~+~$\alpha$~+~$n$ and $^8$Be~+~$n$ continuum states.
The green (dotted) line is that of the resonance pole.
The black thin line represents the sum of them.
The arrow indicates the threshold energy of the $^8$Be(0$^+$)~+~$n$ channel.
}
\end{figure}

We consider the different case when there exists the $1/2^+$ resonance to see how the resonance contributes to the photodisintegaration cross section.
By using the three-body potential with $v_3 = -1.3$ MeV, the $1/2^+$ resonance is obtained with the energy and decay width being 0.091 and 0.002 MeV, respectively, just below the $^8$Be~+~$n$ threshold at 0.0918 MeV measured from the $\alpha$~+~$\alpha$~+~$n$ threshold. 
In FIG.~\ref{fig:deccs_res}, we show the cross section in this case, which shows quite a sharp peak at the resonance energy.
From the decomposition of the cross section, we find that the peak in the cross section is dominated by the resonance while the $^8$Be~+~$n$ continuum states have a sizable contribution to the cross section.
This trend is much different from that in FIG.~\ref{fig:deccs}.
In the result of FIG.~\ref{fig:deccs}, the peak in the cross section is constructed by several low-lying eigenstates located on the $^8$Be~+~$n$ continuum $2\theta$ line in FIG.~\ref{fig:edis}.

{\it Summary.}
We investigate the character of the $1/2^+$ state of $^9$Be using the photodisintegration reaction with the $\alpha$~+~$\alpha$~+~$n$ three-body model and the CSM.
The calculated photodisintegration cross sections into the $1/2^+$ states are shown to have a strong dependence on the strength of the three-body potential for the $1/2^+$ state.
The experimental cross section shows the sharp peak just above the $^8$Be~+~$n$ threshold, which is nicely reproduced with the attractive three-body potential.
We cannot find any resonance poles for the $1/2^+$ states in explaining the peak in the cross section.
From the decomposition of the calculated cross section, it is shown that the $^8$Be~+~$n$ continuum states dominates the cross section to the $1/2^+$ states. 
These results indicates the possibility of the virtual-state nature of the first excited $1/2^+$ state.
In addition, the pole trajectory suggests that the pole of the $1/2^+$ state is located on the second Riemann sheet of $^8$Be~+~$n$ instead of the broad resonances of $\alpha$~+~$\alpha$~+~$n$.


\begin{thebibliography}{99}
\bibitem{Terasawa01} M. Terasawa, K. Sumiyoshi, T. Kajino, G. J. Mathews, and I. Tanihata, Astrophys. J. {\bf 562}, 470, (2001).
\bibitem{Sasaqui05} T. Sasaqui, T. Kajino, G. J. Mathews, K. Otsuki, and T. Nakamura, Astrophys. J. {\bf 634}, 1173 (2005).
\bibitem{Gibbons59} J. H. Gibbons, R. L. Macklin, J. B. Marion, and H. W. Schmitt, Phys. Rev. {\bf 114}, 1319 (1959).
\bibitem{John62} W. John and J. M. Prosser, Phys. Rev. {\bf 127}, 231 (1962).
\bibitem{Fujishiro82} M. Fujishiro, T. Tabata, K. Okamoto, and T. Tsujimoto, Can. J. Phys. {\bf 60}, 1672 (1982).
\bibitem{Utsunomiya00} H. Utsunomiya, Y. Yonezawa, H. Akimune, T. Yamagata, M. Ohta, M. Fujishiro, H. Toyokawa, and H. Ohgaki, Phys. Rev. C {\bf 63}, 018801 (2000).
\bibitem{Burda10} O. Burda, P. von Neumann-Cosel, A. Richter, C. Forss\'en, and B. A. Brown, Phys. Rev. C {\bf 82}, 015808 (2010).
\bibitem{Arnold12} C. W. Arnold, T. B. Clegg, C. Iliadis, H. J. Karwowski, G. C. Rich, J. R. Tompkins, and C. R. Howell, Phys. Rev. C {\bf 85}, 044605 (2012).
\bibitem{Efros99} V. Efros and J. Bang, Eur. Phys. J. A {\bf 4}, 33 (1999).
\bibitem{Arai03} K. Arai, P. Descouvemont, D. Baye, and W. N. Catford, Phys. Rev. C {\bf 68}, 014310 (2003).
\bibitem{Garrido10} E. Garrido, D. Fedorov, and A. Jensen, Phys. Lett. B {\bf 684}, 132 (2010).
\bibitem{A_Rodriguez10} R. \'Alvarez-Rodr\'iguez, A. S. Jensen, E. Garrido, and D. V. Fedorov, Phys. Rev. C {\bf 82}, 034001 (2010).
\bibitem{Garrido11} E. Garrido, R. de Diego, D. Fedorov, and A. Jensen, Eur. Phys. J. A {\bf 47}, 102 (2011), ISSN 1434-6001.
\bibitem{Casal14} J. Casal, M. Rodr\'iguez-Gallardo, J. M. Arias, and I. J. Thompson, Phys. Rev. C {\bf 90}, 044304 (2014).
\bibitem{Aguilar71} J. Aguilar and J. M. Combes, Commun. Math. Phys. {\bf 22}, 269 (1971).
\bibitem{Balslev71} E. Balslev and J. M. Combes, Commun. Math. Phys. {\bf 22}, 280 (1971).
\bibitem{Ho83} Y. K. Ho, Phys. Rep. {\bf 99}, 1 (1983).
\bibitem{Moiseyev98} N. Moiseyev, Phys. Rep. {\bf 302}, 212 (1998).
\bibitem{Aoyama06} S. Aoyama, T. Myo, K. Kat\=o, and K. Ikeda, Prog. Theor. Phys. {\bf 116}, 1 (2006).
\bibitem{Myo14} T. Myo, Y. Kikuchi, H. Masui, and K. Kat\=o, Prog. Part. Nucl. Phys. {\bf 79}, 1 (2014).
\bibitem{Odsuren14} M. Odsuren, K. Kat\=o, M. Aikawa, and T. Myo, Phys. Rev. C {\bf 89}, 034322 (2014).
\bibitem{Myo11b} T. Myo, Y. Kikuchi, and K. Kat\=o, Phys. Rev. C {\bf 84}, 064306 (2011).
\bibitem{Myo12} T. Myo, Y. Kikuchi, and K. Kat\=o, Phys. Rev. C {\bf 85}, 034338 (2012).
\bibitem{Myo01} T. Myo, K. Kat\=o, S. Aoyama, and K. Ikeda, Phys. Rev. C {\bf 63}, 054313 (2001).
\bibitem{Myo08} T. Myo, Y. Kikuchi, K. Kat\=o, H. Toki, and K. Ikeda, Prog. Theor. Phys. {\bf 119}, 561 (2008).
\bibitem{Kikuchi13a} Y. Kikuchi, T. Myo, K. Kat\=o, and K. Ikeda, Phys. Rev. C {\bf 87}, 034606 (2013).
\bibitem{Saito77} S. Saito, Prog. Theor. Phys. Suppl. {\bf 62}, 11 (1977).
\bibitem{Kanada79} H. Kanada, T. Kaneko, S. Nagata, and M. Nomoto, Prog. Theor. Phys. {\bf 61}, 1327 (1979).
\bibitem{Schmid61} E. Schmid and K. Wildermuth, Nuclear Physics {\bf 26}, 463 (1961).
\bibitem{Kukulin86} V. I. Kukulin, V. M. Krasnopol'sky, V. T. Voronchev, and P. B. Sazonov, Nucl. Phys. A {\bf 453}, 365 (1986).
\bibitem{hiyama03} E. Hiyama, Y. Kino, and M. Kamimura, Prog. Part. Nucl. Phys. {\bf 51}, 223 (2003).
\bibitem{Myo97} T. Myo and K. Kat\=o, Prog. Theor. Phys. {\bf 98}, 1275 (1997).
\bibitem{Tilley04} D. Tilley, J. Kelley, J. Godwin, D. Millener, J. Purcell, C. Sheu, and H. Weller, Nucl. Phys. A {\bf 745}, 155 (2004).
\bibitem{Nortershauser09} W. N{\"o}rtersh{\"a}user, D. Tiedemann, M. \v{Z}\'akov\'a, Z. Andjelkovic, K. Blaum, M. L. Bissell, R. Cazan, G. W. F. Drake, C. Geppert, M. Kowalska, et al., Phys. Rev. Lett. {\bf 102}, 062503 (2009).
\bibitem{Tanihata88} I. Tanihata, T. Kobayashi, O. Yamakawa, S. Shimoura, K. Ekuni, K. Sugimoto, N. Takahashi, T. Shimoda, and H. Sato, Phys. Lett. B {\bf 206}, 592 (1988).
\bibitem{Masui00} H. Masui, S. Aoyama, T. Myo, K. Kat\=o, and K. Ikeda, Nucl. Phys. A {\bf 673}, 207 (2000), ISSN 0375-9474.


\end{thebibliography}
\end{document}